# Stoichiometric Tuning of Lattice Flexibility and Na Diffusion in NaAlSiO$_4$: Quasielastic Neutron Scattering Experiment and Ab-initio Molecular Dynamics Simulations

Mayanak K. Gupta[1,2$], Ranjan Mittal[1,3*], Sajan Kumar[1,3], Baltej Singh[1,3], Niina H Jalarvo[5]
Olivier Delaire[2], Rakesh Shukla[4], Srungarpu N. Achary[3,4], Alexander I. Kolesnikov[5],
Avesh K. Tyagi[3,4], and Samrath L. Chaplot[1,3]

[1]*Solid State Physics Division, Bhabha Atomic Research Centre, Mumbai, 400085, India*
[2]*Department of Mechanical Engineering and Materials Science, Duke University, NC USA*
[3]*Homi Bhabha National Institute, Anushaktinagar, Mumbai 400094, India*
[4]*Chemistry Division, Bhabha Atomic Research Centre, Mumbai, 400085, India*
[5]*Neutron Scattering Division, Oak Ridge National Laboratory, Oak Ridge, Tennessee 37831, USA*
Email: mayankg@barc.gov.in[$], rmittal@barc.gov.in[*]

We have performed quasielastic neutron scattering (QENS) experiments up to 1243 K and ab-initio molecular dynamics (AIMD) simulations to investigate the Na diffusion in various phases of NaAlSiO$_4$ (NASO), namely, low-carnegieite (L-NASO; trigonal), high-carnegieite (H-NASO; cubic) and nepheline (N-NASO; hexagonal) phases. The QENS measurements reveal Na ions localized diffusion behavior in L-NASO and N-NASO, but long-range diffusion behavior in H-NASO. The AIMD simulation supplemented the QENS measurements and showed that excess Na ions in H-NASO enhance the host network flexibility and activate the AlO$_4$/SiO$_4$ tetrahedra rotational modes. These framework modes enable the long-range diffusion of Na across a pathway of interstitial sites. The simulations also show Na diffusion in Na-deficient N-NASO through vacant Na sites along the hexagonal *c*-axis.

## I. INTRODUCTION

Li-ion batteries are ubiquitous for energy storage in portable electronics [1-3], yet improving current battery technology demands new materials with higher energy density, longer life cycle, and lower cost [4-6]. Current commercial Li-ion batteries use liquid or gel electrolytes based on organic polymer salts that are flammable and limit the choice of anode materials, in addition to creating issues associated with leakage, chemical instability, vaporization, and dendrite formation [7-9]. The use of solid electrolytes based on superionic conductors [10, 11] could essentially circumvent these issues and widen the range of operating temperature and pressure [10, 11]. Solid electrolytes could enable metallic Li as the anode, further enhancing power density [10, 12]. Such solid-state batteries could offer a most compact and safe version to store energy at various scales [10, 11, 13-15]. However, candidate solid-electrolyte materials exhibit high activation energy for ionic diffusion [11, 13], resulting in a smaller ionic conductivity than liquid or salt-polymer electrolytes [16]. Improving the ionic conductivity of solid electrolytes at room temperature is essential to enable solid-state batteries for applications such as electric vehicles and grid storage [1, 11, 13, 17, 18].

The limited abundance and uneven geographical distribution of lithium in the earth's crust are commercial supply concerns [19, 20]. Therefore, interest in developing alternative Na-based batteries is growing [21-28]. After discovering fast Na conduction in layered β-alumina, many Na-ion conductors such as $Na_3PS_4$, $Na_3SbS_4$, $Na_{11}Sn_2PS_{12}$, $Na_2B_{12}H_{12}$, and $Na_3OBH_4$, have been identified to exhibit very high ionic conductivities (~mS/cm) and attracted attention as candidate solid electrolytes[29-35]. An atomistic investigation of Na diffusion in solid electrolytes and identifying the relevant descriptor would help design better electrolytes. We earlier studied the compound $LiAlSiO_4$ (β-eucryptite), which is fascinating due to its 1-$d$ nature of superionic Li conductivity and nearly zero thermal expansion behavior up to 900 K[36, 37]. We identified the presence of negative thermal expansion (NTE) along the $c$- axis promotes the Li conductivity along the c-axis; hence, NTE could be an important descriptor for materials with similar topological structure.

$NaAlSiO_4$ occurs (**Fig. 1**) in three polymorphic phases, namely, nepheline (N-NASO; hexagonal, $P6_3$), low-carnegieite (L-NASO; trigonal, $P3_2$), and high-carnegieite (H-NASO; cubic, F-43m) phases[38]. X-ray diffraction and TG-DTA data show that L-NASO transforms to H-NASO at about 923 K[38]. L-NASO has an ordered structure with a fully occupied Na at *3a* Wyckoff site[38], while H-NASO ($Na_{1.14}AlSiO_4$) has about 14 % excess sodium with full and partial occupancies at *4d* and *4b* Wyckoff sites. Besides this, O in H-NASO also shows disordered behavior, which leads to many possible orientations of $AlO_4$ and $SiO_4$ tetrahedral units and causing a flexible network. Further, in N-NASO ($Na_{0.873}AlSiO_4$), Na occupies the *2a* and *6c* Wyckoff sites with ~49% and 100% occupancies, respectively. Both L-NASO and N-NASO phases are stable at room temperature[38]. A high-temperature Raman spectroscopy study shows a significant change in the Raman spectra at 400 cm$^{-1}$ near the L-NASO to H-NASO transition[39]. Also, the ionic conductivity measurements are reported in the L-NASO but limited to 500K (~10$^{-5}$ mS/cm at 500 K) [40].

QENS is a powerful technique to probe the stochastic dynamics of atoms and molecules in liquid and solids. It measures the elastic broadening caused by random collisions between the neutron and diffusive elements in the sample. However, QENS studies on Na/Li-based ionic conductors are sparse due to the low neutron scattering cross-section of these elements. A few successful QENS studies have investigated Na-ion diffusion, notably in sodium cobaltate [41], amorphous $Na_2Si_2O_5$ [42], $Na_x[Ni_{1/3}Ti_{2/3}]O_2$ [43], and $Na_3SbS_4$ [30]. Earlier, we successfully exploited QENS and AIMD techniques [36, 37] to investigate various Li and Na-based solid ionic conductors to understand the diffusion mechanism and thermodynamic stability in respective compounds[36, 37, 42, 44]. Here we present QENS measurements and AIMD simulations to investigate Na-stoichiometry impact on host dynamics and Na diffusion. The QENS measurements show that only H-NASO exhibits long-range Na diffusion, while the other two phases show localized Na diffusion. Hence, understanding the Na diffusion mechanism in different phases would help us identify possible new descriptors for Na diffusion in such compounds and help engineer the materials for energy storage applications.

## II. EXPERIMENTAL

The N-NASO phase was prepared by heating stochiometric amounts of $Na_2CO_3$, $\gamma$-$Al_2O_3$, and $SiO_2$ (quartz) with an excess of 5 mol % $Na_2CO_3$ in a procedure as reported earlier [45]. The pellet of the homogenous mixture of the reactant was heated at several progressively increasing temperatures, viz. 823 K, 12h, 1023 K, 12 h, 1123 K, 12 h and finally at 1223 K, 12h. After heating at each temperature, the pellet was crushed to powder, homogenized, pelletized, and used for the subsequent heating. Further, the powder sample of N-NASO was pressed to cylindrical pellets of 20 mm diameter and 20 mm height and then heated at 1623 K for 6h, followed by cooling to obtain the L-NASO phase. Further heating of L-NASO above 923 K transforms to the H-NASO phase. We used a heating and cooling rate ~2 K/minute for sample synthesis, and different phases were characterized using X-ray diffraction.

The QENS experiments were performed using the BASIS time-of-flight backscattering spectrometer at the Spallation Neutron Source at Oak Ridge National Laboratory, USA[46]. For these measurements, we have used 5.38 and 6.13 grams of L-NASO and N-NASO polycrystalline samples, respectively, in cylindrical annular platinum sample holders of 25 mm diameter and 2 mm gap for the sample. We have performed the experiments at 300 K, 500 K, 700 K, 900 K, 1100 K, and 1243 K. The samples were heated with MICAS furnace with quartz tube insert for purging dry air (to avoid oxygen release at high temperatures) available at ORNL[47]. The elastic energy of 2.08 meV was chosen for the measurements with Si(111) analyzers, giving an overall energy resolution ~ 3.7 μeV. The QENS data were collected in the $Q$-range from 0.2 to 2.0 Å$^{-1}$, with an accessible range of energy transfer from −100 to 100 μeV. The data analysis was performed using the Mantid[48] and Dave[49] software packages. The data were fitted with a delta function (due to elastic scattering), one Lorentzian function (due to QENS), and a linear background convoluted with the instrumental resolution. The room temperature data were used as the resolution function as all diffusive motions were essentially frozen at this temperature for the samples studied in this work.

## III. COMPUTATION DETAILS

The VASP software [50] was used to perform AIMD simulations to understand the diffusion of Na atoms in various phases of $NaAlSiO_4$ up to 1400 K. The AIMD simulations were performed with a time step of 2 femtoseconds using a single k-point at the zone center in the Brillouin zone, and 900 eV energy cut-off for the plane-wave expansion. An energy convergence criterion of 10$^{-6}$ eV was chosen for self-consistent energy convergence. The generalized gradient approximation (GGA) exchange-correlation parametrized by Perdew, Burke, and Ernzerhof within projected–augmented wave formalism was used[51,52].

The AIMD simulations were performed at 300 K, 800 K, 900 K, 1000 K, 1200 K, 1300 K, and 1400 K with NVT ensemble, where the Nose thermostat represents the heat bath. The first 5 ps of the trajectories, during which the simulation reached equilibrium, were discarded to evaluate the thermodynamical and transport properties. The lattice parameters of the different NASO polymorphs at 300 K are: N-NASO (a=9.976 Å, c=8.350 Å, hexagonal); L-NASO ($a$ = 5.109 Å, $c$ = 12.488 Å, hexagonal); and H-NASO (a= 7.733 Å, cubic). We used a supercell of dimension $\sqrt{2} \times \sqrt{2} \times 2$, $3 \times 3 \times 1$, and $2 \times 2 \times 2$ for N-NASO, L-NASO, and H-NASO, respectively, which contains 220, 189, and 229 atoms, respectively.

## IV. RESULTS AND DISCUSSION
### A. Quasielastic Neutron Scattering

In **Fig S1 and S2** (Supplementary Material[53]), we showed the temperature-dependent QENS spectra in different phases of NASO. We observed an apparent gradual increase of quasielastic broadening with increasing temperature. The QENS spectra were fit with the sum of a delta function (elastic line), a Lorentzian (quasielastic spectrum), and a linear background, convoluted with the instrument's resolution function. The half-width at half-maximum (HWHM) of the Lorentzian component provides direct information about diffusion characteristics. Long-range diffusion is characterized by a $Q^2$ dependence of HWHM in the low $Q$ regime, while localized diffusion (e.g., atoms or molecules in pores/ cages) shows a nearly constant HWHM as a function of $Q$.

NASO consists of the cage-like structure formed by $AlO_4/SiO_4$ units filled by Na atoms. The Na diffusion across these cages depends on the cage's permeability (flexibility of host lattice to allow the Na migration). Here we show that the host flexibility can be tuned by changing the Na-stochiometry, hence the Na- diffusion. We will describe in a latter section that the presence of excess Na in NASO enhances the Na permeability by flexing the host framework and activating the re-orientational dynamics. In contrast, we find that presence of Na vacancy is less effective in promoting long-range diffusion. In **Figure 2(b)**, we have shown the $Q$ dependence of QENS estimated HWHM across a range of temperatures. The L-NASO phase at 700 K does not show any significant $Q$ dependence of HWHM, which refers to a localized Na dynamics within $AlO_4/SiO_4$ cages. At 900 K and above, we start observing the $Q$-dependence of HWHM at the low-$Q$ regime. This infers the onset of long-range Na diffusion above 900 K. To estimate the Na diffusion constant above 900 K, we used the Chudley-Elliot jump-diffusion model to analyze the $Q$ dependence of HWHM as given by[54]:

$$\Gamma(Q)=(1-\mathrm{Sin}(Qd)/Qd)/\tau$$

Here $d$ is the jump length of diffusing ions and $\tau$ is the residence time. The estimated values of $d$ and $\tau$ obtained from fitting are given in TABLE-I. It may be noted that at above 923 K, the L-NASO is known to transform to H-NASO [38]; hence, it is quite possible that at 900 K, the L-NASO sample partially transforms to H-NASO even below 923 K and exhibits the Na diffusion. The jump length obtained from the C-E model in the L-NASO phase at 900 K is 2.9(1) Å. While at 1100 K (H-NASO), the estimated jump length is ~ 4.0 Å. The calculated pair distribution function (g(r)) of H-NASO shows the first neighbor Na-Na distance spanning from 2.5 to 4.1 Å (**Fig 3**). So the QENS estimated jump lengths above 900 K lie within the range of first-neighbor Na-Na distances in the H-NASO phase and correspond to the distance between excess Na at the interstitial site (Na1) with the regular site (Na2). Hence, the QENS measurements infer that the long-range diffusion in H-NASO primarily occurs through interstitial sites and are supported by our AIMD simulations. In contrast, N-NASO does not show any significant $Q$ dependence of HWHM up to 1243 K, which implies the absence of long-range Na diffusion. It seems that interstitial site presence is necessary to perform the inter-cage Na hops, which is forbidden in L-NASO and N-NASO phases. We elaborate more on it in the next section. Our QENS data reveal a long-range diffusion in the H-NASO phase in contrast to N-NASO and L-NASO. The estimated diffusion coefficient in H-NASO is ~$10^{-9}$ m$^2$/sec, similar to that reported in other Na-based superionic compounds with layered structure from QENS measurements[30, 41, 43].

**B. Molecular Dynamics Investigation**

To further investigate Na diffusion and underlying mechanism at the microscopic level, we have performed the AIMD simulation in all three phases of NASO from 300 K to 1400 K.

**Figure 3** and **Figure S5** (Supplementary Material[53]) show the calculated time-averaged pair-distribution function g(r) between Na-Na and Na-host-elements (O, Al, and Si) using AIMD simulations. The g(r) between Na and host elements shows sharp peaks in L-NASO, while in H-NASO, it's slightly broad and more apparent in Na-O pairs. Besides this, Na-Na g(r) in H-NASO shows an additional peak at ~3.4 Å, which appears due to excess Na at interstitial sites. In comparison, Na-Na g(r) in N-NASO shows a shallow peak at ~4.7 Å, indicating a broad distribution of Na-Na first neighbors with sizeable thermal amplitude of Na within the polyhedral cages.

The calculated MSD of O, Na, Al, and Si atoms in all three phases of NASO at elevated temperatures is shown in **Fig 4 (a) and Fig. S6 (**Supplementary Material[53]**)**. We do not observe Na diffusion in the L-NASO phase below 900 K. The calculated MSD of each species in L-NASO oscillates about some mean value and do not increase during the 60 ps simulation time. At ~ 900 K, we find that few Na atoms show intermittent large MSD values corresponding to a jump-length ~3 Å (**Fig 5(a)**); however, this jump in MSD does not show up for a long-time and resets to a previous value, which does not give any diffusion. This indicates that the jump attempts in the L-NASO phase quash by the polyhedral network. Hence, it seems that Na remains localized within their respective AlO4/SiO4 cages, which is also reflected in the QENS measurements of the L-NASO phase.

In the case of H-NASO, we see a nearly linear increase in Na MSD with time (**Fig. 4(a)** and **Fig. S6(b),** Supplementary Material[53]), which confirms the presence of Na diffusion. The presence of extra Na in H-NASO plays a vital role in the long-range Na diffusion onset. It acts in two ways to promote the diffusion, (i) enhanced Coulombic repulsion between Na's reduces the diffusion barrier, and (ii) creates distortion in the cage structure, which offers more flexibility for AlO$_4$ and SiO$_4$ dynamics. This essentially opens up the gateways for the Na migration between cages through interstitial sites and generates the 3-$d$ channels of Na diffusion in H-NASO. The estimated diffusion constant (**Fig. S7,** Supplementary Material[53]) from MSD is much smaller than that from QENS measurements. The underestimated diffusion constant in AIMD could be due to limited AIMD trajectory and different Na stoichiometry in the measured sample. However, the simulations qualitatively explain the observed long-range diffusion in QENS measurements. We also performed AIMD simulation in the H-NASO phase without interstitial Na (excess Na) at *4b* sites at 1000 K and 1300 K (**Fig. S8,** Supplementary Material[53]). We did not observe any sign of long-range diffusion up to 20 ps, which establishes that the excess Na atoms in the lattice facilitates the Na diffusion in initiating the diffusion process at low temperatures in H-NASO.

As described before, N-NASO is a Na deficient phase, which also has Na vacant sites in the structure. Often the vacancies enhance the diffusion behavior [55-57] as the diffusing ions may quickly jump through unoccupied sites. Hence, they are one of the viable options to enhance the diffusion in crystalline materials. Na in N-NASO (hexagonal P6$_3$ space group) partially occupies the *2a* (Na1) sites and fully occupies the *6c* (Na2) Wyckoff sites. From AIMD simulation, we find that only Na1 shows jumps along the *c*-axis, while Na2 remains localized at their site **(Fig S6(d-e),** Supplementary Material[53]). The Na1-Na1 site distance in N-NASO is *c*/2 = 4.2 Å. The calculated squared displacement of a few Na1 atoms in the N-NASO at 1300 K is shown in **Fig 5** and **Fig. S10** (Supplementary Material[53]). We observed most of the jump lengths correspond to ~4.5 Å and their multiple. The calculated MSD infers that up to 1100 K (**Fig S6,** Supplementary Material[53]), Na is not exhibiting the long-range diffusion and hoping within neighboring sites along the c-axis leads to a localized diffusion. While at 1300 K and above, it starts diffusing along the $c$-axis (**Fig S6(e),** Supplementary Material[53]). However, due to restricted 1-$d$ diffusion behavior, the bulk diffusion constant in N-NASO is smaller (**Fig. 4(a)**) than in H-NASO. Hence, Na vacancies in the N-NASO phase promote Na diffusion but to a much lesser extent than excess Na in H-NASO. In **Fig 5,** we have shown the trajectory of a few selected Na atoms in all three NASO phases, which depicts the Na localized diffusion in L-NASO, long-range 3-$d$ diffusion in H-NASO, and 1-$d$ diffusion in N-NASO.

As discussed above, the different Na stoichiometry plays a critical role in promoting long-range diffusion. It affects the interaction between Na-Na as well as the host structure topology to reduce the migration barrier. To illustrate both the effects, we have calculated (i) the time dependence of bond angles between Al-O bonds and the c-axis of crystal and (ii) the Na probability iso-surface plot, shown in **Figs 4(b) and 4(c),** respectively in all three phases.

In L-NASO, the time-dependent angles between Al-O/Si-O bonds and the c-axis fluctuate about the mean values and do not indicate any re-orientation or anomalous magnitude of rotation of the $AlO_4$ polyhedral units. We observe similar behavior of bond-angles in N-NASO. However, in the H-NASO phase, the bond angles show significant fluctuation and infer the re-orientation of $AlO_4$ units and a very flexible cage structure. The calculated probability density iso-surface plot of Na occupation in three phases of NASO show the Na hopping pathways. It established that the Na diffusion in the H-NASO phase occurs via the hopping between regular and interstitial Na sites and does not involve other pathways. Importantly $AlO_4/SiO_4$ librational and re-orientational dynamics in H-NASO enable the Na hoping.

Further, in the L-NASO phase, Na are arranged in 1-$d$ channels and primarily localized in $AlO_4/SiO_4$ cages. While in N-NASO, Na vacancy reduces the Coulomb barrier and creates local distortion, which enhances the probability of inter-cage hopping along the c-axis. Still, the long-range diffusion is limited to a small value as compared with H-NASO. Overall the analysis indicates that the excess Na doping is a promising approach in enhancing the ionic conductivity in NASO and likely in such corner shared framework structures.

## V. CONCLUSIONS

QENS measurements along with AIMD simulations are used to understand the Na diffusion in different phases of $NaAlSiO_4$. The presence of excess Na in H-NASO creates the dynamical frustration in $AlO_4$/$SiO_4$ to activate the paddle-wheel mechanism for long-range Na diffusion in H-NASO. The presence of smaller Na-Na distance reduces Na migration's barrier energy and enhances the Na diffusion in H-NASO. The L-NASO and Na-deficient N-NASO phase show very limited diffusion behavior due to localized Na dynamics in polyhedral cages. Further, excess Na atoms in H-NASO generate 3-$d$ diffusion channels compared to 1-$d$ channels in N-NASO, enhancing the long-range diffusion. Our study based on QENS and AIMD simulations suggests that in such framework structure compounds, by performing stochiometric engineering, one can tune the diffusion properties by flexing the host lattice and activating the paddle-wheel mechanism.

**ACKNOWLEDGEMENTS**

The use of the ANUPAM super-computing facility at BARC is acknowledged. SLC acknowledges the financial support of the Indian National Science Academy for the INSA Senior Scientist position. A portion of this research used resources at the Spallation Neutron Source, a DOE Office of Science User Facility operated by the Oak Ridge National Laboratory.

TABLE-I The jump length (d) and a jump time ($\tau$) as obtained from the fitting of CE model[54] to the Q-dependent variation of half-width-at-half-maximum (HWHM) of Lorentzian peak extracted from dynamic neutron scattering function S(Q, E) of the L-NASO phase (900 K) and H-NASO phase of $NaAlSiO_4$. For L-NASO phase at 700 K, and N-NASO phase, only the jump time is fitted using a localized diffusion model. L-NASO is known to transform [38] to H-NASO phase at 923 K. (1 μeV$^{-1}$=658 ps)

|  | T (K) | CE Model | | | | Localized diffusion Model | |
|---|---|---|---|---|---|---|---|
|  |  | d (Å) | $\tau$ (μeV$^{-1}$) | $\tau$ (ps) | D (× 10$^{-10}$ m$^2$/sec) | $\tau$ (μeV$^{-1}$) | $\tau$ (ps) |
| L-NASO | 700 |  |  |  |  | 0.01513±0.0003 | 10.0±0.2 |
|  | 900 | 2.9±0.1 | 0.02072 ±0.0004 | 13.6±0.3 | 10±1 |  |  |
| H-NASO | 1100 | 4.0±0.2 | 0.01368±0.0007 | 9.0±0.5 | 30±5 |  |  |
|  | 1243 | 4.2±0.3 | 0.01279±0.0008 | 8.4±0.5 | 35±7 |  |  |
| N-NASO | 700 |  |  |  |  | 0.01307±0.0004 | 8.6±0.3 |
|  | 900 |  |  |  |  | 0.01384±0.0004 | 9.1±0.3 |
|  | 1100 |  |  |  |  | 0.01384±0.0003 | 9.1±0.2 |
|  | 1243 |  |  |  |  | 0.01343±0.0001 | 8.8±0.1 |

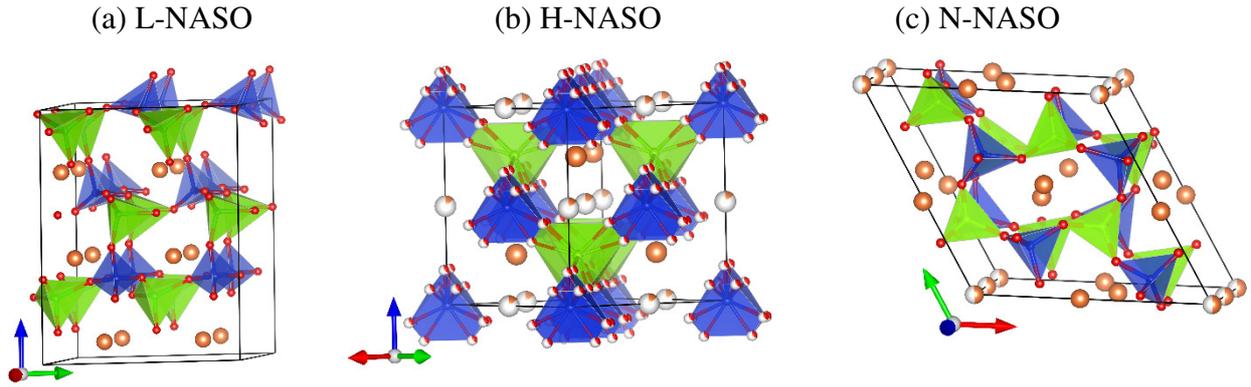

FIG 1 (Color online) Structure of L-NASO, H-NASO, and N-NASO. The color scheme is Na: Orange, Al: Green, Si: Blue and O: Red; $AlO_4$: Green and $SiO_4$: Blue. Na partial occupancy in H-NASO and N-NASO phases are shown by partially orange sphere. The crystallographic axis is shown by red, green, and blue arrows at the bottom representing the *a-*, *b-,* and *c-* axis, respectively. L-NASO structure is shown in a (2×2×1) supercell, while the other two phases are shown in the conventional unit cell.

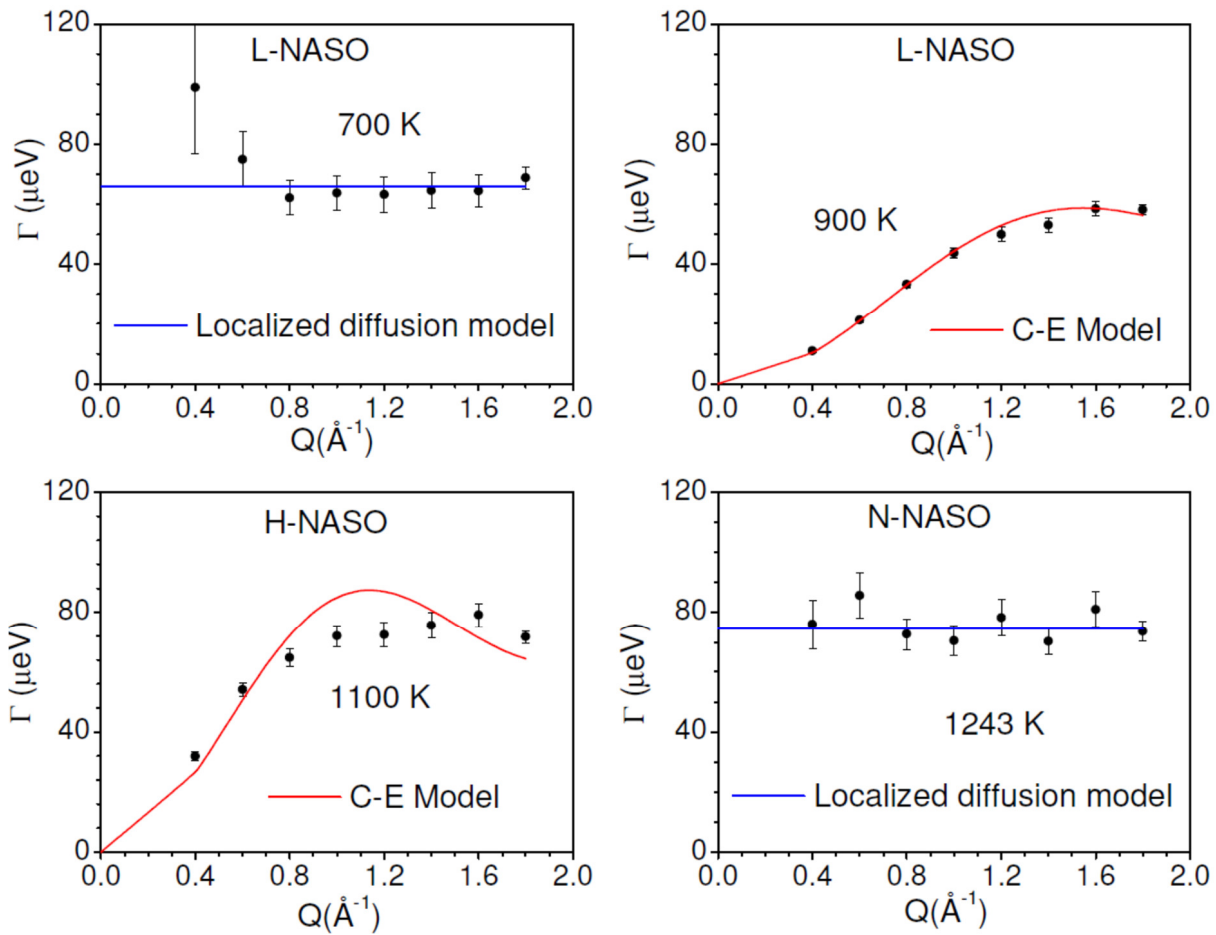

FIG 2 (Color online) The estimated HWHM of QENS spectra in the different phases of NASO. (a) L-NASO, 700K (b) L-NASO at 900K, (c) H-NASO at 1100 K, and (d) N-NASO phase at 1243 K, respectively. L-NASO is known to transform to H-NASO phase at 923 K[38].

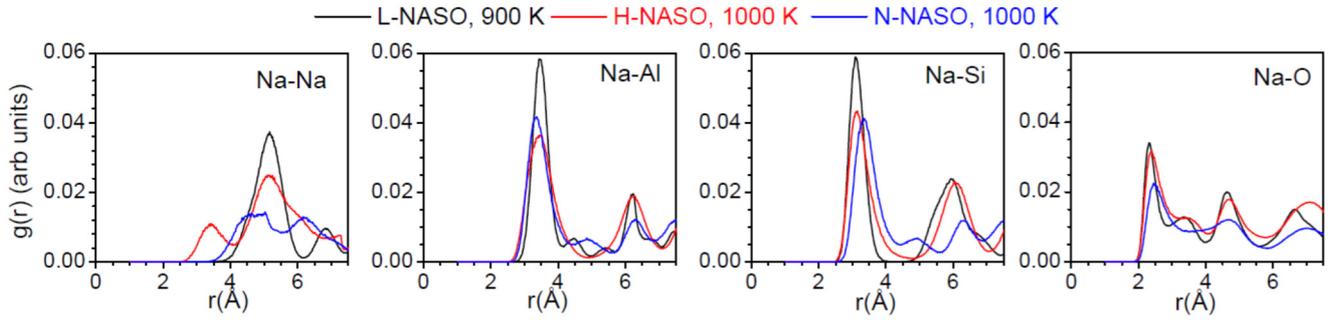

FIG 3 (Color online) The calculated pair distribution functions of different pairs of atoms in the low-carnegieite (L-NASO; trigonal, $P3_2$), high-carnegieite (H-NASO; cubic, F-43m), and nepheline (N-NASO; hexagonal, $P6_3$) phase of $NaAlSiO_4$.

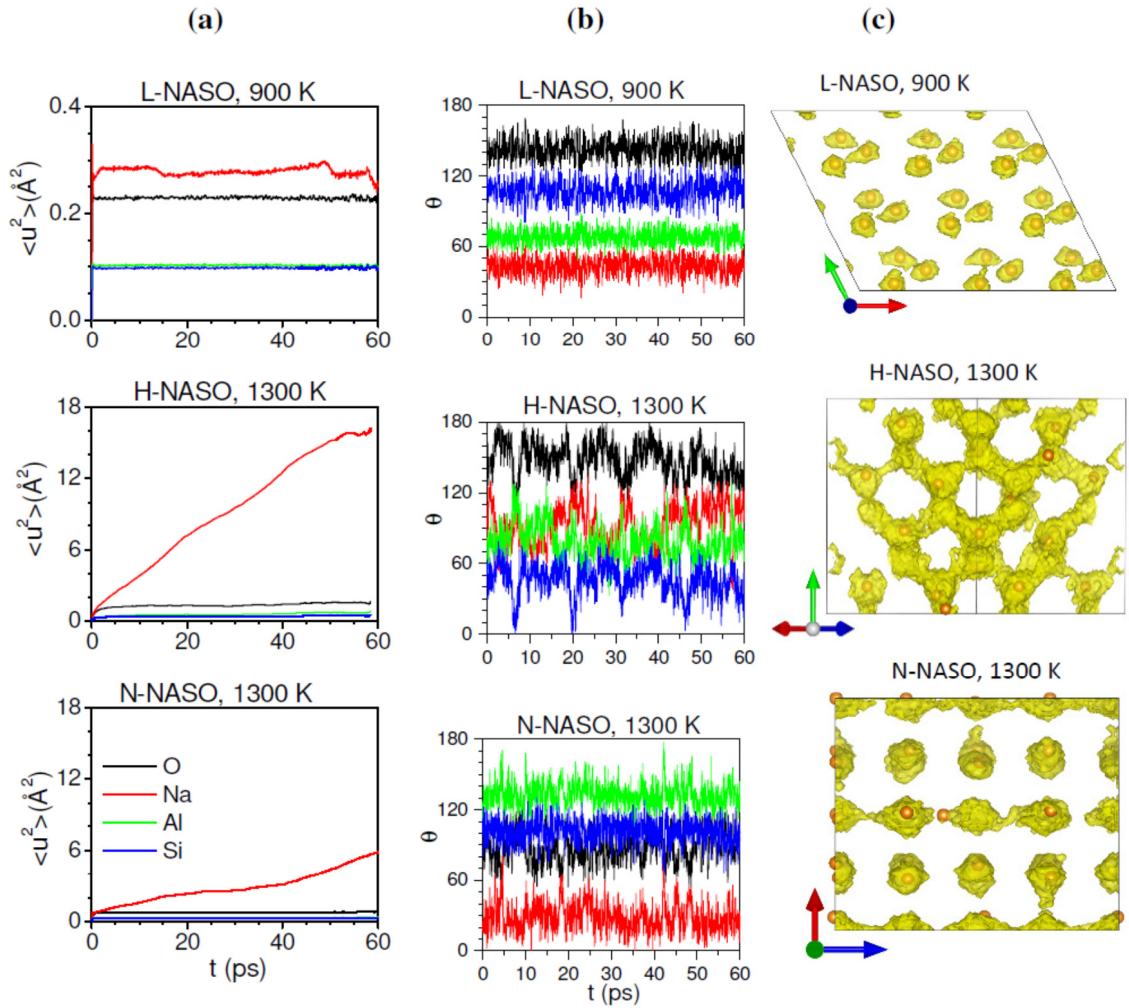

FIG 4 (Color online) (a) The calculated MSD of Na in different NASO phases. (b) The calculated time dependence of angle between Al-O bonds and c- axis of a representative $AlO_4$ tetrahedral unit. (c) The probability iso-surface density of Na atoms. The orange spheres show the equilibrium positions of Na. The crystallographic axis is shown by red, green, and blue arrows at the bottom representing the *a*-, *b*-, and *c*- axis, respectively.

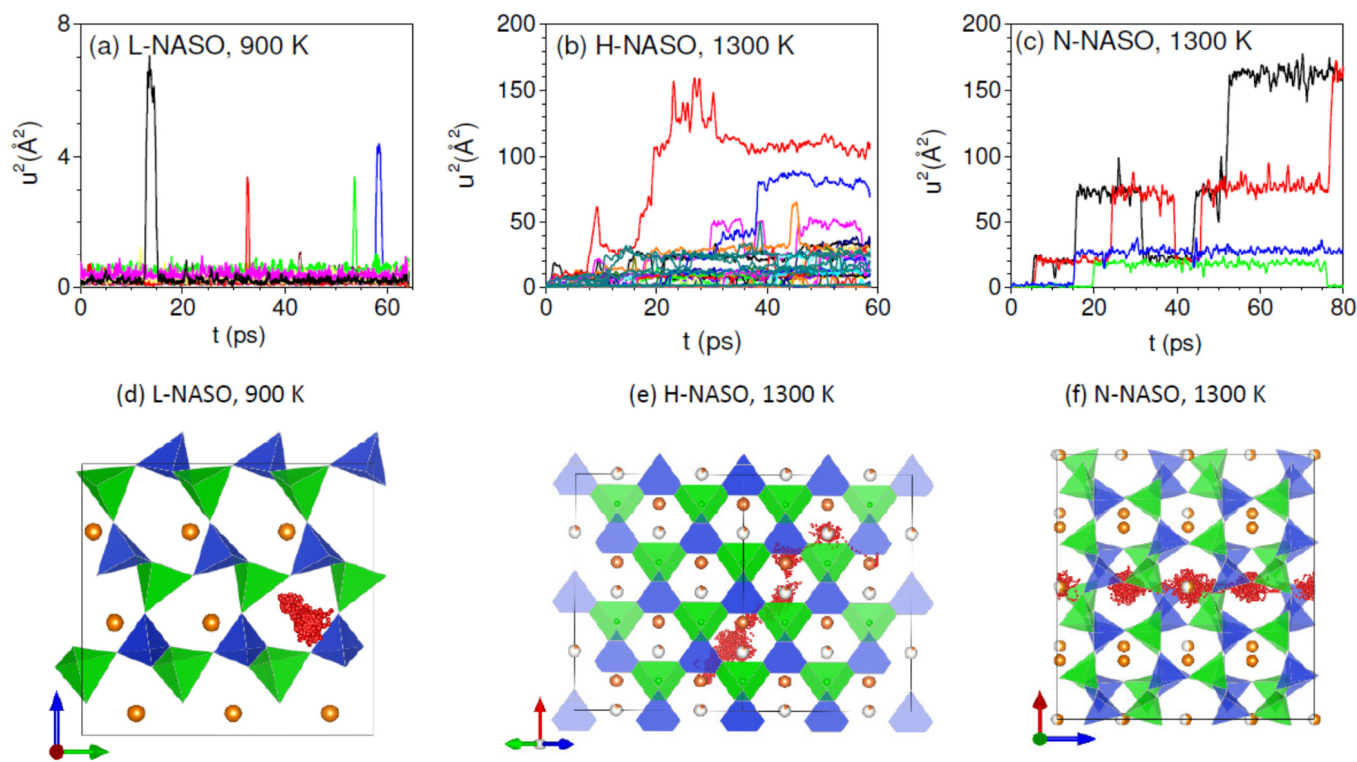

FIG 5 (Color online) The calculated squared displacement of a few representatives Na in (a) L-NASO, (b) H-NASO, and (c) N-NASO. A representative Na atom trajectory (red dots) in the framework structure of (d) L-NASO exhibits localized diffusion, (e) H-NASO shows the 3-$d$ diffusion, and (f) N-NASO shows 1-$d$ Na diffusion. Blue and green polyhedral units show the $AlO_4$ and $SiO_4$ units, and orange spheres show the equilibrium Na position. Na partial occupancy in H-NASO and N-NASO phases are shown by partially orange sphere. The crystallographic axis is shown by red, green, and blue arrows at the bottom representing the $a$-, $b$-, and $c$- axis, respectively.